\documentclass[twocolumn,aps,showpacs,amsmath,floatfix,superscriptaddress]{revtex4}
\usepackage{graphicx}
\usepackage{dcolumn}
\usepackage{bm}
\usepackage{hyperref} 
\usepackage{latexsym}
\usepackage{float}

\begin{document}
\title{Random walk with shrinking steps}
\author{P.~L.~Krapivsky}
\email{paulk@bu.edu}
\author{S.~Redner}
\email{redner@bu.edu}
\affiliation{Center for BioDynamics, Center for Polymer Studies, 
and Department of Physics, Boston University, Boston, Massachusetts 02215}

\begin{abstract}
  
  We outline the properties of a symmetric random walk in one dimension in
  which the length of the $n$th step equals $\lambda^n$, with $\lambda<1$. As
  the number of steps $N\to\infty$, the probability that the endpoint is at
  $x$ approaches a limiting distribution $P_\lambda(x)$ with many beautiful
  features. For $\lambda<1/2$, the support of $P_\lambda(x)$ is a Cantor set.
  For $1/2\leq \lambda<1$, there is a countably infinite set of $\lambda$
  values for which $P_\lambda(x)$ is singular, while $P_\lambda(x)$ is smooth
  for almost all other $\lambda$ values. In the most interesting case of
  $\lambda= g \equiv (\sqrt{5}-1)/2$, $P_g(x)$ is riddled with singularities
  and is strikingly self-similar. The self-similarity is exploited to derive
  a simple form for the probability measure $M(a,b)\equiv\!\int_a^b
  P_g(x)\,dx$.

\end{abstract}


\maketitle

\section{INTRODUCTION}

This article discusses the properties of random walks in one dimension in
which the length of the $n$ step changes systematically with $n$. That is,
at the $n$th step, a particle hops either to the right or to the left with
equal probability by a distance $f(n)$, where $f(n)$ is a monotonic but
otherwise arbitrary function. The usual nearest-neighbor random walk is
recovered when $f(n)=1$ for all $n$.

Why should we care about random walks with variable step lengths? There are
several reasons. First, for the {\em geometric random walk}, where
$f(n)=\lambda^n$ with $\lambda<1$, a variety of beautiful and unanticipated
features arise,\cite{W,E,G,Kac} as illustrated in Fig.~\ref{pdf}. A very
surprising feature is that the character of the probability distribution of
the walk changes dramatically as $\lambda$ is changed by a small amount.
Most of our discussion will focus on this intriguing aspect. We will 
emphasize the case
$\lambda=g \equiv (\sqrt{5}-1)/2$, the inverse of the golden ratio, where
the probability distribution has a beautiful self-similar appearance. We will
show how many important features of this distribution can be obtained by
exploiting the self-similarity, as well as the unique numerical properties of
$g$.

There also are a variety of unexpected applications of random walks with
variable step lengths. One example is spectral line broadening in
single-molecule spectroscopy.\cite{BS} Here the energy of a chromophore in a
disordered solid reflects the interactions between the chromophore and the
molecules of the host solid. For a dipolar interaction potential and for
two-state host molecules with intrinsic energies $\pm \epsilon$, the energy
of the chromophore is proportional to $\sum_j \pm \epsilon r_{j}^{-3}$, where
$r_{j}$ is the separation between the chromophore and the $j$th molecule.
This energy is equivalent to the displacement of a geometric random walk
with $f(n)= n^{-3}$. 

Another example is the motion of a Brownian particle in a fluid with a linear
shear flow, that is, the velocity field is $v_x(y)=Ay$. As a Brownian
particle makes progressively longer excursions in the $\pm y$ directions
(proportional to $t^{1/2}$), the particle experiences correspondingly larger
velocities in the $x$-direction (also proportional to $t^{1/2}$). This gives
rise to an effective random walk process in the longitudinal direction in
which the mean length of the $n$th step grows linearly with $n$, that is,
$f(n)=n$.\cite{BRA}

Historically, the geometric random walk has been discussed, mostly in the
mathematics literature, starting in the 1930's.\cite{W,E} Interest in such
random walks has recently revived because of connections with dynamical
systems.\cite{A,L} Recent reviews of the geometric random walk can be found
in Ref.~\onlinecite{PSS}; see also Ref.~\onlinecite{DF} for a review of more
general iterated random maps. In contrast, there appears to be no mention of
the geometric random walk in the physics literature, aside from
Ref.~\cite{junk}.

Finally, the geometric random walk provides an instructive set of examples
that can be analyzed by classical probability theory and statistical physics
tools.\cite{reif,weiss} These examples can serve as a useful pedagogical
supplement to a student's introduction to the theory of random walks. As we
shall discuss, there are specific values of $\lambda$ for which the
probability distribution can be calculated by elementary methods, while for
other $\lambda$ values, there is meager progress toward an exact solution,
even though many tantalizing clues exist.

\section{Picture Gallery}

The displacement of a one-dimensional random walk after $N+1$ steps has the
form $x_N= \sum_{n=0}^N \epsilon_n f(n)$, where each $\epsilon_n$ takes on
the values $\pm 1$ with equal probability. Consequently, the mean-square
displacement $\langle x_N^2\rangle$ can be expressed as $\sum_n f(n)^2$. If
this sum is finite, then there is a finite mean-square displacement as
$N\to\infty$. In such a case, the endpoint probability distribution therefore
approaches a fixed limit.\cite{bazant} Henceforth, we will focus on the case
of geometrically shrinking step lengths, that is, $f(n)=\lambda^n$ with
$\lambda<1$.  We denote the endpoint probability distribution after $N$ steps
by $P_\lambda(x,N)$ and limiting form $\lim_{N\to\infty} P_\lambda(x,N)$ by
$P_\lambda(x)$.  We will show that $P_\lambda(x)$ exhibits rich behavior as
$\lambda$ is varied.

\noindent
 \includegraphics*[width=0.40\textwidth]{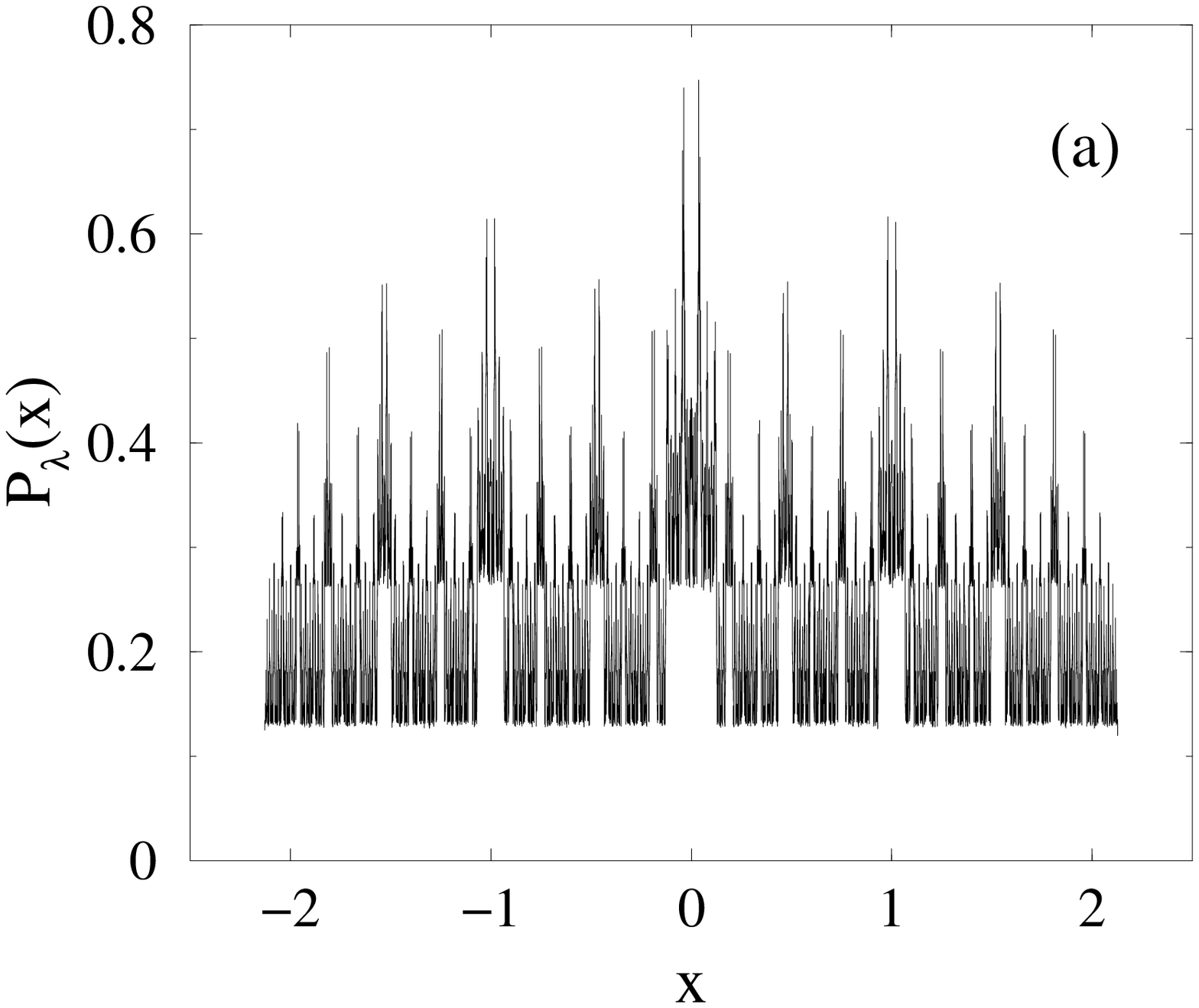} \\
 \includegraphics*[width=0.40\textwidth]{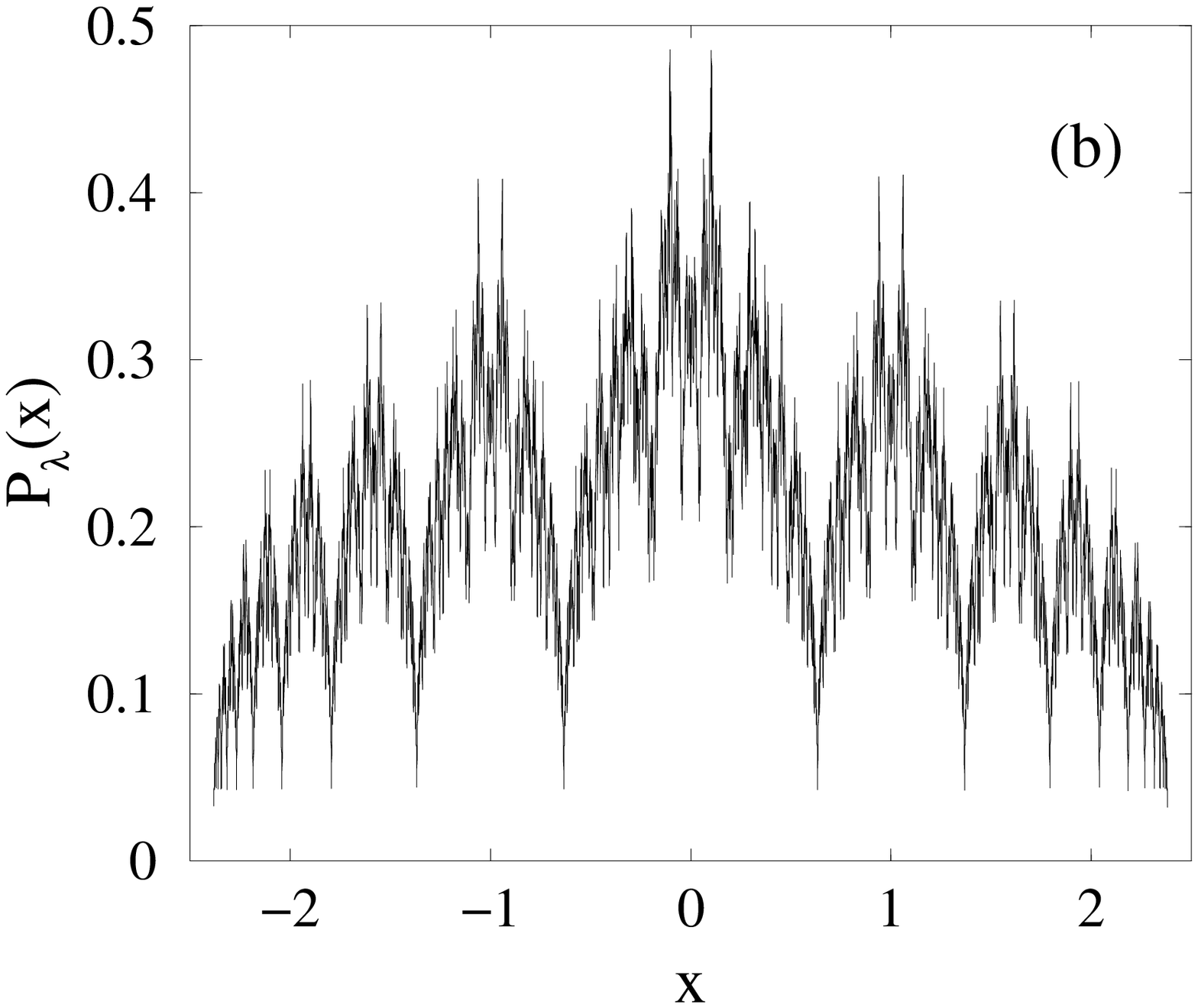} \\
 \includegraphics*[width=0.40\textwidth]{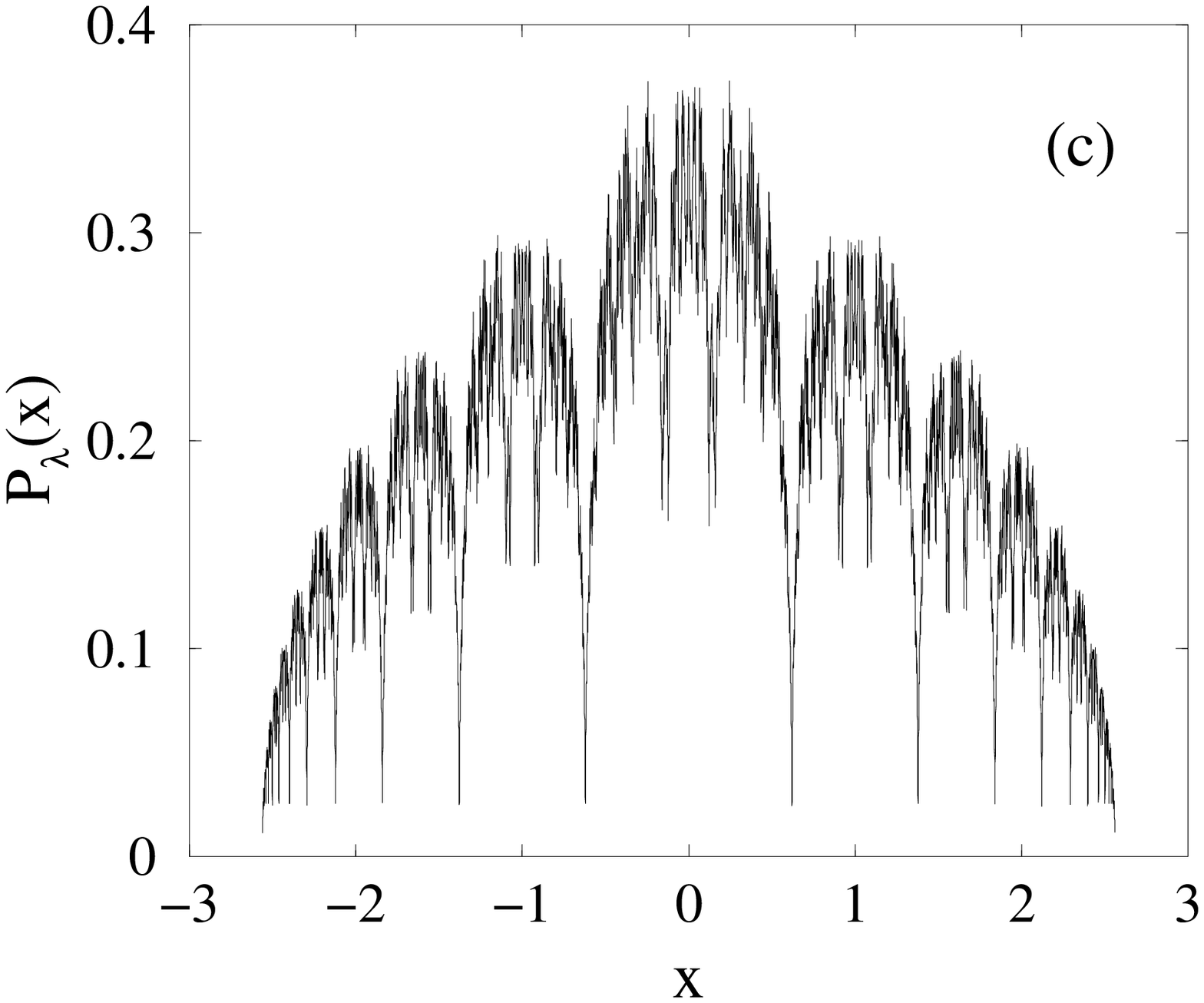} \\
 \includegraphics*[width=0.40\textwidth]{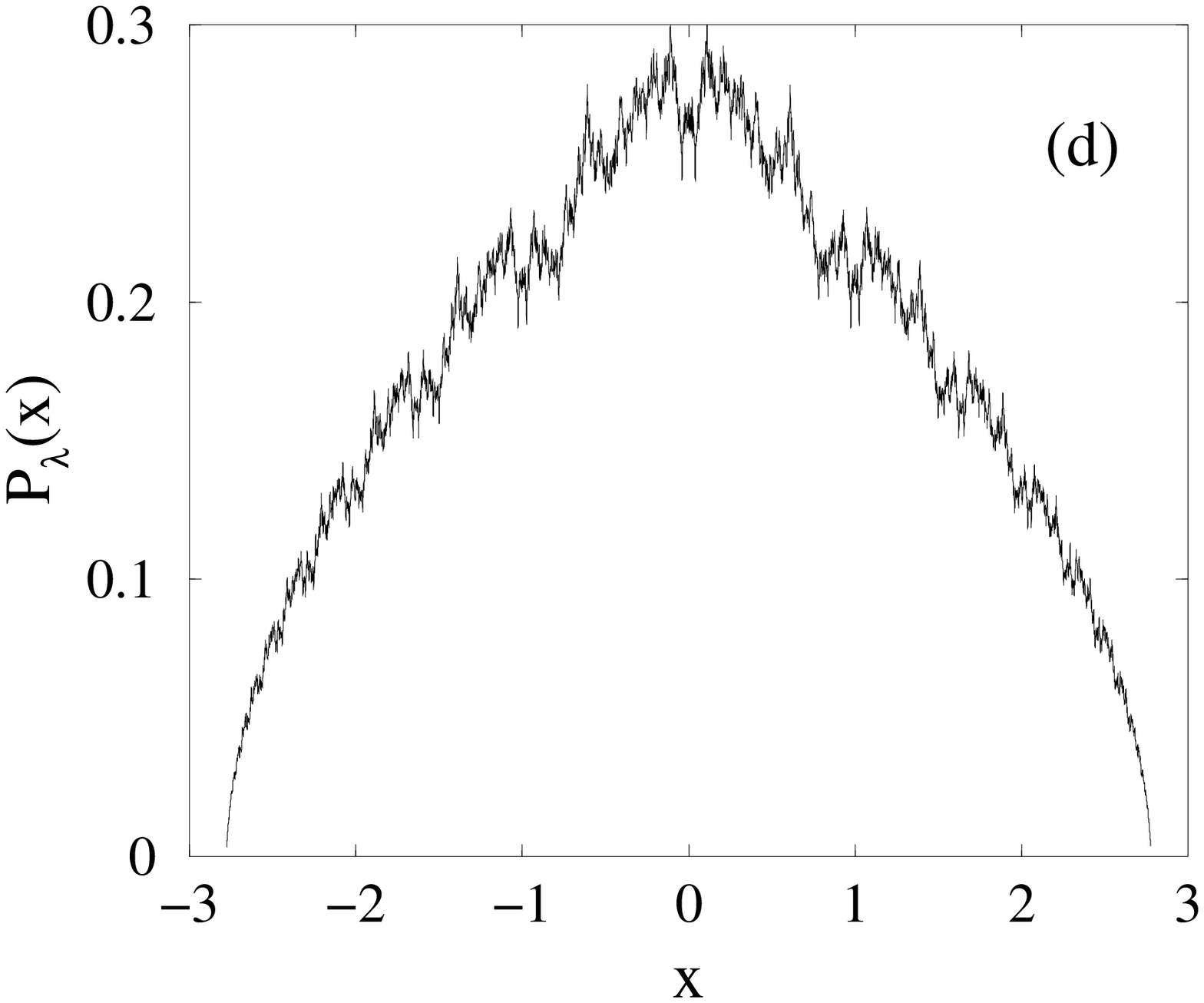} \\
 \includegraphics*[width=0.40\textwidth]{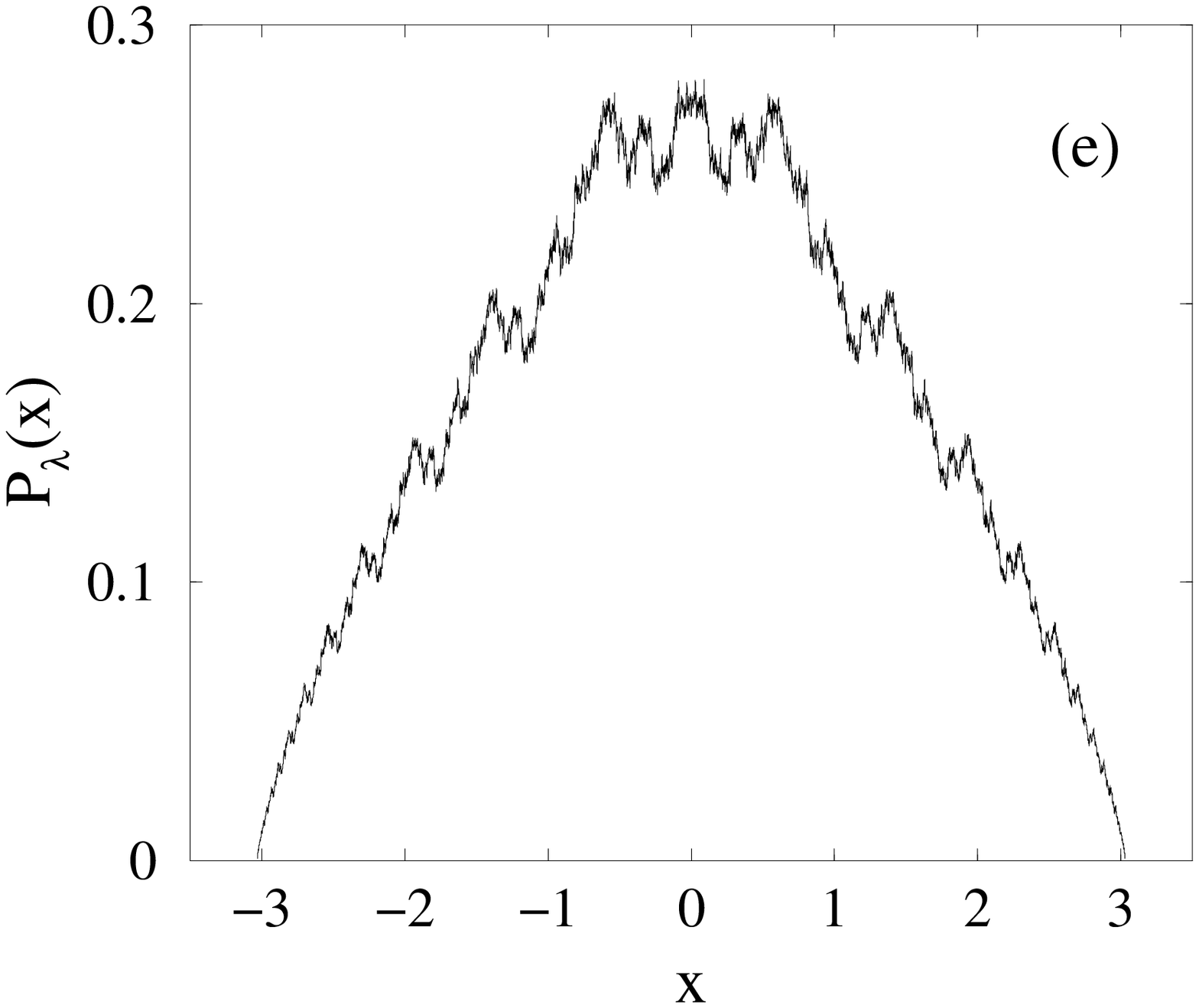} \\
 \includegraphics*[width=0.40\textwidth]{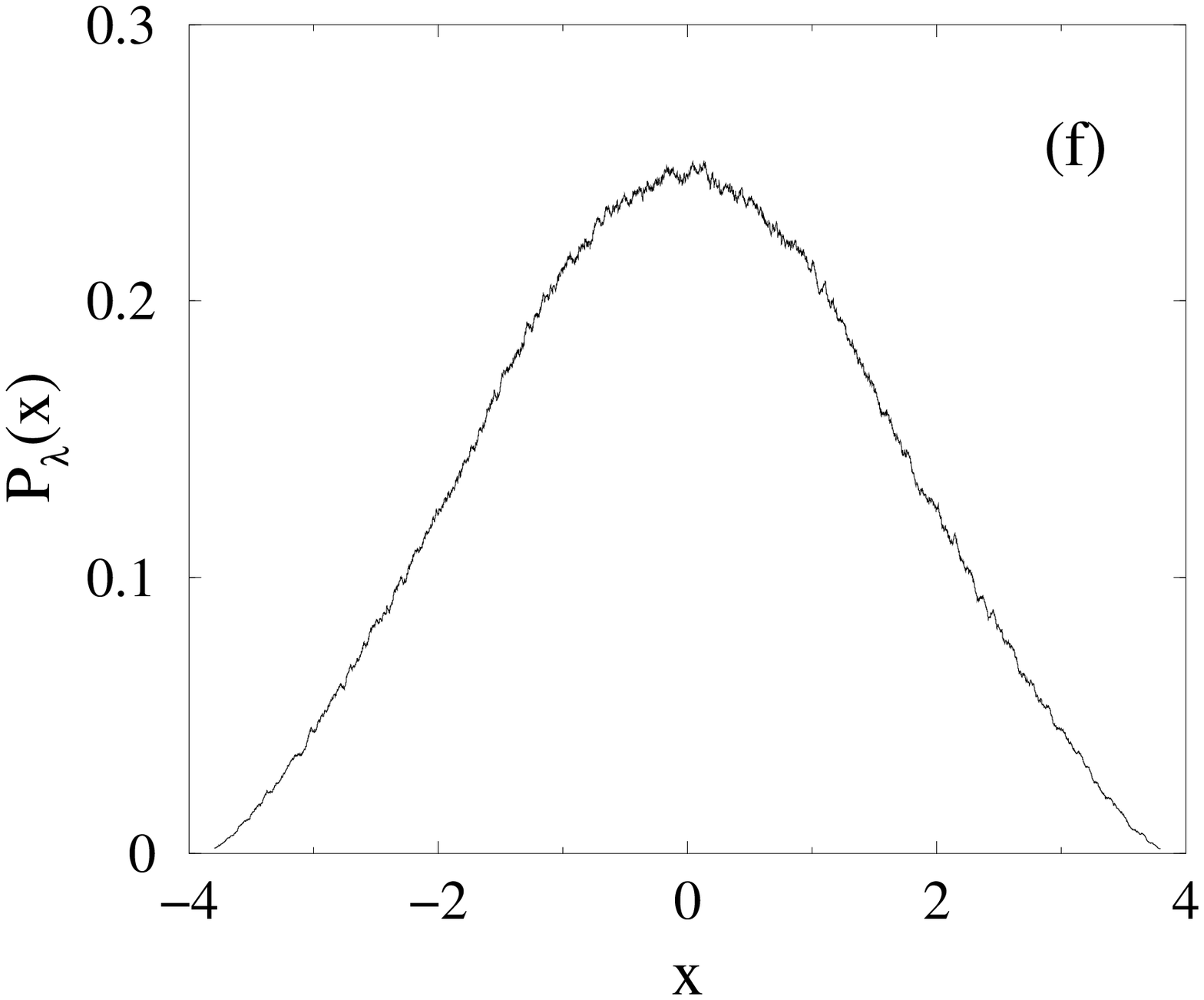} \\
\vskip -3ex
\begin{figure}[H]
\caption{Simulated probability distribution $P_\lambda(x)$ for 
  $\lambda = 0.53, 0.58, 0.61, 0.64, 0.67$, and 0.74 ((a) -- (f)).  The data
  for each $\lambda$ is based on $10^8$ realizations of 40 steps at spatial
  resolution $10^{-3}$.
\label{pdf}}
\end{figure}

To obtain a qualitative impression of $P_\lambda(x)$, a small picture gallery
of $P_\lambda(x)$ for representative values of $\lambda>1/2$ is given in
Fig.~\ref{pdf}. As we shall discuss, when $\lambda=1/2$, $P_{1/2}(x)=1/4$ for
$|x|\leq 2$ and $P_{1/2}(x)=0$ otherwise.  For $\lambda<1/2$, the support of
the distribution is fractal (see below).  As $\lambda$ increases from 1/2 to
approximately 0.61, $P_\lambda(x)$ develops a spiky appearance that changes
qualitatively from multiple local maxima to multiple local minima (see
Fig.~\ref{pdf}). In spite of this singular appearance, it has been proven by
Solomyak\cite{S} that the cumulative distribution is absolutely continuous
for almost all $\lambda>1/2$. On the other hand, Erd\H{o}s\cite{E} showed
that there is a countably infinite set of $\lambda$ values, given by the
reciprocal of the Pisot numbers in the range $(1,2)$, for which
$P_\lambda(x)$ is singular.  A Pisot number is an algebraic number (a root of
a polynomial with integer coefficients) all of whose conjugates (the other
roots of the polynomial) are less than one in modulus (more details about
Pisot numbers can be found in Ref.\cite{pisot}).  It is still unknown,
however, if these Pisot numbers constitute all of the possible $\lambda$
values for which $P_\lambda(x)$ is singular.  For $\lambda>0.61$,
$P_\lambda(x)$ rapidly smooths out and beyond $\lambda \agt 0.7$, there is
little visual evidence of spikiness in the distribution at the $10^{-3}$
resolution scale of Fig.~\ref{pdf}.

There is a simple subset of $\lambda$ values for which singular behavior can
be expected on intuitive grounds. In particular, consider the situations
where $\lambda$ satisfies
\begin{equation}
\label{condition}
1-\sum_{n=1}^N \lambda^n=0.\nonumber
\end{equation}
This condition can be viewed geometrically as a walk whose first step (of
length 1) is to the right and $N$ subsequent steps are to the left such that
the walker returns {\em exactly} to the origin after $N+1$ steps.  This
positional degeneracy, in which points are reached by different walks with
the same number of steps, appears to underlie the singularities in
$P_\lambda(x)$.  The roots of Eq.~(\ref{condition}) give the solution
$\lambda=g \equiv (\sqrt{5}-1)/2\approx 0.618$, 0.5437, 0.5188, 0.5087,
0.5041 \ldots for $N=2,3,\ldots$ As we shall discuss, the largest in this
sequence, $\lambda= g$ (the inverse of the golden ratio), leads to especially
appealing behavior and the distribution $P_g(x)$ has a beautiful
self-similarity as well as an infinite set of
singularities.\cite{E,G,A,LP,SV}

\section{General Features of the Probability Distribution}

\begin{figure}[ht] 
 \vspace*{0.cm}
 \includegraphics*[width=0.4\textwidth]{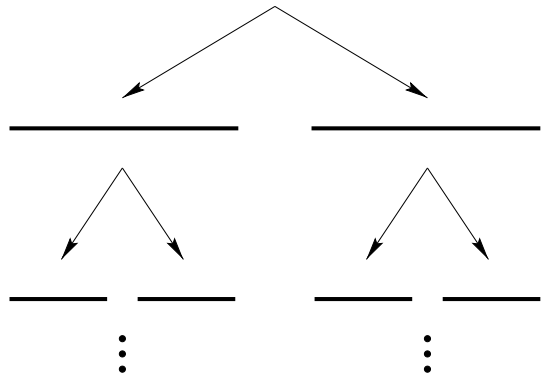}
\caption{Fragmentation of the support of the probability distribution during the 
  first two steps of the shrinking random walk when $\lambda<1/2$.  After $k$
  steps, the support fragments into $2^k$ intervals each of length
  $2\lambda^k/(1-\lambda)$.
\label{cantor}}
\end{figure}

For $\lambda<1/2$, the support of $P_\lambda(x)$, namely the subset of the
real line where the distribution is non-zero, is a Cantor set. To understand
how this phenomenon arises, suppose that the first step is to the right. The
maximum displacement of the subsequent walk is $\lambda/(1-\lambda)$.
Consequently, the endpoint of the random walk necessarily lies within the
region $\big[1-\frac{\lambda}{1-\lambda},1+\frac{\lambda}{1-\lambda}\big]$.
Thus the support of $P_\lambda(x)$ divides into two non-overlapping regions
after one step (see Fig.~\ref{cantor}). This same type of bifurcation occurs
at each step, but at a progressively finer distance scale so that the support
of $P_\lambda(x)$ breaks up into a Cantor set.

The evolution of the support of the probability distribution can be
determined in a precise way by recasting the geometric random walk into the
random map,
\begin{equation}
\label{map}
x'=\pm 1 +\lambda x,
\end{equation}
which describes how the position of a particle changes in a single step. We
can see that this map is equivalent to the original random walk process by
substituting $x'$ for $x$ on the right-hand side of Eq.~(\ref{map}) and
iterating.  Therefore the map is equivalent to the random sum
$x=\sum_n^\infty \epsilon_n\lambda^n$.

Similarly, the equation for the probability distribution at the
$N$th step satisfies the recursion relation\cite{reif,weiss}
\begin{equation}
\label{basic}
P_\lambda(x,N)=\frac{1}{2}\Big[P_\lambda
\big(\frac{x-1}{\lambda},N-1\big)+
P_\lambda\big(\frac{x+1}{\lambda},N-1\big)\Big].
\end{equation} Equation~(\ref{basic}) states that for a particle to be at
$x$ at the
$N$th step, the particle must either have been either at $(x+1)/\lambda$ or
at
$(x-1)/\lambda$ at the previous step; the particle then hops to $x$ with
probability 1/2. For any
$\lambda<1$, the probability distribution necessarily approaches a fixed
limit $P_\lambda(x)$ for $N\to\infty$. Consequently, $P_\lambda(x)$ remains
invariant under the mapping given by Eq.~(\ref{map}) and thus satisfies the
invariance condition
\begin{equation}
\label{Px}
P_\lambda(x)=\frac{1}{2}
\Big[P_\lambda\big(\frac{x-1}{\lambda}\big)+
P_\lambda\big(\frac{x+1}{\lambda}\big)\Big].
\end{equation}

Because $P_\lambda(x)$ can have a singular appearance, it often is more
useful to characterize this distribution by the {\em probability measure}
$M_\lambda(a,b)$ defined by
\begin{equation}
\label{nu}
M_\lambda(a,b)=\!\int_a^b \! dx\,P_\lambda(x).
\end{equation}
The integral in Eq.~(\ref{nu}) smooths out singularities in $P_\lambda$
itself, and we shall see that it is more amenable to theoretical analysis.
The invariance condition of Eq.~(\ref{Px}) can be rewritten in
terms of this measure as
\begin{eqnarray}
\label{M}
2M_\lambda(a,b)=M_\lambda \Big(\frac{a - 1}{\lambda}, 
\frac{b - 1}{\lambda}\Big) + 
M_\lambda \Big(\frac{a + 1}{\lambda}, 
\frac{b + 1}{\lambda}\Big).
\end{eqnarray}

Equation~(\ref{M}) can now be used to determine the support of $M_\lambda$.
Clearly, the support of $M_\lambda$ lies within the interval
$J_\lambda=[-x_{\max},x_{\max}]$, with $x_{\max}=1/(1-\lambda)$.
For $\lambda<1/2$, the map (\ref{map}) transforms the interval $J_\lambda$
into the union of the two non-overlapping subintervals (see
Fig.~\ref{cantor}),
\begin{equation}
\label{sub}
\Big[-\frac{1}{(1-\lambda)},-
\frac{(1-2\lambda)}{(1-\lambda)}\Big],\
\Big[\frac{(1-2\lambda)}{(1-\lambda)},
\frac{1}{(1-\lambda)}\Big].
\end{equation}
By restricting the map (\ref{map}) to these two subintervals, we find
that they are transformed into four non-overlapping subintervals after
another iteration. If we continue these iterations {\em ad infinitum}, we
obtain a support for $M_\lambda$ that consists of a collection of
disjoint sets that ultimately comprise a Cantor set.\cite{W}

On the other hand, for $\lambda\geq 1/2$, the map again transforms
$J_\lambda$ into the two subintervals given in Eq.~(\ref{sub}), but now these
subintervals are overlapping. Thus the support of $P_\lambda$ fills the
entire range $[-x_{\max},x_{\max}]$.

\section{Exact Distribution for $\lambda=2^{-1/m}$}

In this section, we derive $P_\lambda$ by Fourier transform methods for
$\lambda=2^{-1/m}$. We will illustrate that the different values of $m$ turn
out to be exactly soluble because of a set of fortuitous cancellations in the
product form for the Fourier transform of the probability distribution.

For a general random walk process, the probability $P(x,N)$ that the endpoint
of the walk is at $x$ at the $(N+1)$ step obeys the fundamental
convolution equation\cite{reif,weiss}
\begin{equation}
\label{convol}
P(x,N)=\sum_{x'} P(x-x',N-1) \,p_N(x'),
\end{equation} 
where $p_N(x)$ is the probability of hopping a distance $x$ at the $N^{\rm
  th}$ step.  Equation~(\ref{convol}) expresses the fact that to reach $x$
after $(N+1)$ steps the walk must first reach a neighboring point $x-x'$
after $N$ steps and then hop from $x-x'$ to $x$ at step $(N+1)$. The
convolution structure of Eq.~(\ref{convol}) cries out for employing Fourier
transforms.  Thus we introduce
\begin{eqnarray}
\label{ftdef}
p_N(k)&=&\!\int_{-\infty}^\infty\! p_N(x)\, e^{ikx}\, dx \nonumber \\
P(k,N)&=&\!\int_{-\infty}^\infty\! P(x,N)\, e^{ikx}\, dx,
\end{eqnarray}
and substitute these forms into Eq.~(\ref{convol}). If the
random walk is restricted to integer-valued lattice points, these
integrals become discrete sums. The Fourier transform turns the convolution
in $x$ into a product in $k$-space,\cite{arfken} and therefore
Eq.~(\ref{convol}) becomes the recursion relation
\begin{equation}
\label{recur}
P(k,N)=P(k,N-1)\,p_N(k).
\end{equation}

We now iterate Eq.~(\ref{recur}) to obtain the formal solution
\begin{equation}
\label{formal} P(k,N)=P(k,0)\,\prod_{n=0}^N p_n(k).
\end{equation}
Generally, we consider the situation where the random walk
begins at the origin. Thus $P(x,0)=\delta_{x,0}$ and correspondingly from
the second of Eq.~(\ref{ftdef}), $P(k,0)=1$. To
calculate
$P(x,N)$, we evaluate the product in Eq.~(\ref{formal}) and then invert the
Fourier transform.\cite{reif} To simplify the notation for the examples of
this section, we define
$\Pi_m(x)= P_{2^{-1/m}}(x)$. We explicitly consider the cases $m=1$, 2,
and 3. From these results, the qualitative behavior for general $m$ is then
easy to understand. 

The single-step probability distribution at the $n$th step is
\begin{equation}
\label{pn}
p_n(x)=\frac{1}{2}[\delta(x-\lambda^n)+\delta(x+\lambda^n)], \nonumber
\end{equation}
and the corresponding Fourier transform is 
\begin{eqnarray}
\label{ft}
p_n(k)&=&\!\int_{-\infty}^\infty p_n(x) e^{ikx}\, dx =\cos(k\lambda^n).
\end{eqnarray}
The Fourier transform of the probability distribution after $N+1$ steps is
the product of the Fourier transforms of the single-step
probabilities.\cite{reif,weiss} Thus from Eqs.~(\ref{formal}) and (\ref{ft})
we have
\begin{equation}
\label{prod-formal}
P_\lambda(k,N)=\prod_{n=0}^N \cos(k\lambda^n).
\end{equation}

We now apply this exact solution to the illustrative cases of $\lambda
=2^{1/m}$. For the simplest case of $\lambda=2^{-1}$, the step length
systematically decreases by a factor of 2. By enumerating all walks of a
small number of steps, it is easy to see that the probability distribution is
uniformly distributed on a periodic lattice whose spacing shrinks by a factor
of two at each step. This is a precursor of the uniform nature of
$\Pi_{1}(x)$ in the $N\to\infty$ limit. Algebraically, the product in
Eq.~(\ref{prod-formal}) can be simplified by using the trigonometric
half-angle formula to yield
\begin{eqnarray}
\label{Pk}
\Pi_1(k,N)& = &\cos k\,\, \cos(k/2)\,\ldots\,
\cos(k/2^N) \nonumber \\
& = & \frac{\sin(2k)}{2\sin k} \,\, \frac{\sin k}{2\sin(k/2)} \ldots
\frac{\sin(k/2^{N-1})}{2\sin(k/2^N)} \nonumber \\
& = & \frac{\sin(2k)}{2^{N+1}\sin(k/2^N)} \nonumber\\ 
& \to &\frac{\sin(2k)}{2k}\ \mbox{as}\ N\to\infty.
\end{eqnarray}
Thus the inverse Fourier transform gives an amazingly simple result. The
probability distribution is merely a square-wave pulse, with $\Pi_{1}=1/4$ on
the interval $[-2,2]$ and $\Pi_{1}=0$ otherwise.

The distribution for $\lambda=2^{-1/2}$ can be calculated similarly.  The
successive cancellation of adjacent numerators and denominators as in 
Eq.~(\ref{Pk}) still occurs.  These cancellations become more apparent by
separating the factors that involve $\sin(k/2^j)$ and $\sin(k/2^{j+1/2})$.
Then by following exactly the same steps as those leading to Eq.~(\ref{Pk}),
we obtain the Fourier transform
\begin{equation}
\label{prod}
\Pi_{2}(k)= 
\frac{\sin(2k)}{2k} \frac{\sin(\sqrt{2} k)} {\sqrt{2}k}.
\end{equation}
This product form has a simple interpretation in real space. If we partition
the walk into odd steps $(1,3,5,\ldots)$ and even steps $(2,4,6,\ldots)$,
then both the odd and even steps are separately geometrical random walks with
$\lambda=2^{-1}$, but with the initial step length of the odd walk equal to
$1$ and that of the even walk equal to $1/\sqrt{2}$. In real space, the full
probability distribution for $\lambda=2^{-1/2}$ is just the convolutions of
the constituent distributions for these odd and even walks. Thus in Fourier
space, the full distribution is just the product of the constituent
distributions, as given in Eq.~(\ref{prod}).

\begin{figure}[ht] 
 \vspace*{0.cm}
 \includegraphics*[width=0.4\textwidth]{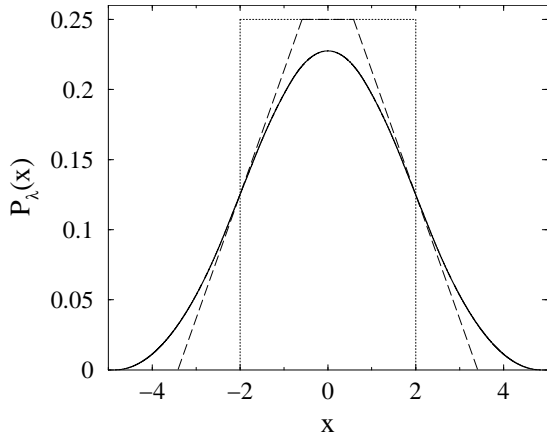}
\caption{Probability distributions for the GW for $\lambda=2^{-1/m}$ for $m=1$,
  2, and 3 (dotted, dashed, and solid, respectively).
\label{2m}}
\end{figure}

To invert the Fourier transform in Eq.~(\ref{prod}) by a direct approach is
straightforward but unwieldy, and the details are given in Appendix A.
Another approach is to use the fact that the probability distribution is the
convolution of two rectangular-shaped distributions -- one in the range
$[-2,2]$ for the odd-step walks and the other in $[-\sqrt{2},\sqrt{2}]$ for
the even-step walks. Thus
\begin{equation}
\label{c2}
\Pi_{2}(x)=\!\int_{-\infty}^\infty\! dx'\,\Pi_{1}(x')\times 
\sqrt{2}\Pi_{1}\Big(\frac{x-x'}{\sqrt{2}}\Big).
\end{equation}
Either by this direct approach or by straightforward Fourier inversion as
given in the Appendix, the final result is (Fig.~\ref{2m})
\begin{eqnarray}
\Pi_{2}(x)=
\begin{cases}
\frac{1}{4} & \text{($|x|<2-\sqrt{2}$)} \\
\frac{1}{4}\big(1-\frac{|x|}{{2+\sqrt{2}}}\big) &
\text{$(2-\sqrt{2}<|x|<2+\sqrt{2})$} \\
0 & \text{$(|x|>2+\sqrt{2})$}
\end{cases}
\end{eqnarray}
Thus the distribution is continuous, but its first derivative is
discontinuous at the four points $x=\pm 2\pm\sqrt{2}$.

Continuing this same train of logic, the solution for general
$\lambda=2^{-1/m}$ is
\begin{equation}
\Pi_{m}(k)=\frac{\prod_{j=1}^m\sin(2^{j/m}k)}{2^{(m+1)/2}k^m}.
\end{equation}
For example, for $\lambda=2^{-1/3}$, the resulting probability
distribution in real space is
\begin{eqnarray*}
\Pi_{3}(x)=
\begin{cases}
\frac{1}{64}(x_4^2-x_3^2-x_2^2-x_1^2-x^2) & (|x|<x_1) \\
\frac{1}{64}(x_4^2-x_3^2-x_2^2-x^2-2xx_1) & (x_1<|x|<x_2) \\
\frac{1}{64}(x_4^2-x_3^2-2x(x_1+x_2)) & (x_2<|x|<x_3) \\
\frac{1}{64}(x-x_4)^2 & (x_3<|x|<x_4) \\
0 & (|x|>x_4)
\end{cases}
\end{eqnarray*}
where
\begin{subequations}
\begin{eqnarray*}
x_1 &=& -2+2^{2/3}+2^{1/3}\approx 0.08473 \\
x_2 &=& +2-2^{2/3}+2^{1/3}\approx 1.6725 \\
x_3 &=& +2+2^{2/3}-2^{1/3}\approx 2.3275 \\
x_4 &=& +2+2^{2/3}+2^{1/3}\approx 4.8473 
\end{eqnarray*}
\end{subequations}
This distribution contains both linear and quadratic segments such that the
first two derivatives of $\Pi_{3}$ are continuous, but the second derivative
is discontinuous at the joining points $x_j$, for $j=1$, 2, 3, 4
(Fig.~\ref{2m}). Generally, for $\lambda=2^{-1/m}$, the distribution has
continuous derivatives up to order $m-1$, while the $m$th derivative is
discontinuous at $2m$ points. As $m\to\infty$, the distribution
progressively becomes smoother and ultimately approaches the Gaussian of the
nearest-neighbor random walk.

A final note about the Fourier transform method is that it provides a
convenient way to calculate the moments,
\begin{equation}
\label{mom}
\langle x^{2k}\rangle =\!\int x^{2k} P_\lambda(x)\, dx,
\end{equation}
for all values of $\lambda$.\cite{vanK} By expanding $P_\lambda(k)$ is a
power series for small $k$, we have
\begin{eqnarray}
\label{mom-f}
P_\lambda(k)&=&\!\int\! dx\, P_\lambda(x)\, e^{ikx}\nonumber \\
&=& \!\int\! dx\, P_\lambda(x)\,
(1+ikx-k^2x^2/2!+\ldots)\nonumber
\\
&=& 1 -\frac{k^2\langle x^2\rangle}{2!} +\frac{k^4\langle x^4\rangle}{4!}+\ldots.
\end{eqnarray}
That is, the Fourier transform contains all the moments of the distribution.
For this reason, the Fourier transform is often termed the moment
generating function.

We take the Fourier transform of the probability distribution of the
geometric random walk and expand this expression in a power series in $k$ to
give
\begin{eqnarray}
\label{exp-f}
P_\lambda(k)&=&\cos k \, \cos(\lambda k)\, \cos(\lambda^2k)\ldots\nonumber \\
&=& \big[1\!-\!\frac{k^2}{2!}\!+\!\ldots\big]
\big[1\!-\!\frac{(\lambda k)^2}{2!}\!+\!\ldots\big]
 \big[1\!-\!\frac{(\lambda^2 k)^2}{2!}\!+\!\ldots\big]\! \ldots\!\nonumber \\
&=& 1 - \frac{k^2}{2}(1+\lambda^2+\lambda^4+\ldots)+{\cal O}(k^4).\nonumber
\end{eqnarray}
If we equate the two power series (\ref{mom-f}) and (\ref{exp-f}) term by
term, we obtain
\begin{subequations}
\label{moments}
\begin{eqnarray}
\langle{x^2}\rangle&=&\frac{1}{1-\lambda^2} \\
\langle{x^4}\rangle&=&\frac{1}{1-\lambda^4}\Big(1+
\frac{6\lambda^2}{1-\lambda^2}\Big).
\end{eqnarray}
\end{subequations}
Moments of any order can be obtained by this approach.

\section{Golden walk}
Particularly beautiful behavior of $P_\lambda(x)$ occurs when $\lambda=g$
(see Fig.~\ref{pdf}(d)). Unfortunately, a straightforward simulation of the
geometric random walk is not a practical way to visualize fine-scale details
of this probability distribution accurately because the resolution is
necessarily limited by the width of the bin used to store the distribution.
We now describe an enumeration approach that is exact up to the number of
steps in the walk. 

\subsection{Enumeration}

It is simple to specify all walks of a given number of steps. Each walk can
be represented as a string of binary digits with 0 representing a step to
the left and 1 representing a step to the right. Thus we merely need to list
all possible binary numbers with a given number of digits $N$ to enumerate
all $N$-step walks. However, we need a method that provides the endpoint
location without any loss of accuracy to resolve the fine details of the
probability distribution.

\begin{figure}[ht] 
 \vspace*{0.cm}
 \includegraphics*[width=0.4\textwidth]{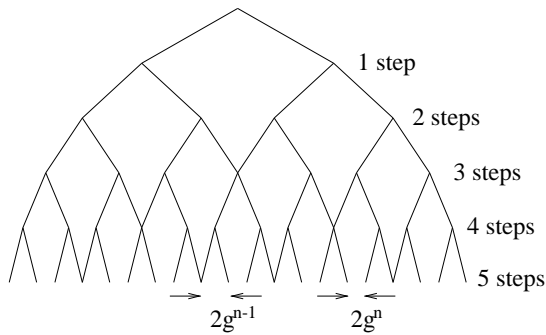}
\caption{First 5 steps of the golden walk enumeration tree.  Notice that the
  distance between adjacent endpoints can only be either $2g^n$ or $2g^{n-1}$.
\label{tree}}
\end{figure}

The basic problem of determining the endpoint locations may be illustrated by
enumerating the first few steps of the walk and representing them as the
branches of a tree as shown in Fig.~\ref{tree}. For
$\lambda=g$, neighboring branches of the tree may rejoin because of the
existence of 3-step walks (one step right followed by two steps left or vice
versa) that form closed loops. For large $N$, the accuracy in the position
of a walk is necessarily lost by roundoff errors if we attempt to evaluate
the sum for the endpoint location, $X_N=\sum_n^N\epsilon_n g^n$, directly.
Thus the recombination of branches in the tree will eventually be missed,
leading to fine-scale inaccuracy in the probability distribution.

However, we may take advantage of the algebra of the golden ratio to reduce
the $N$th-order polynomial in $X_N$ to a first-order polynomial. To
this end, we successively use the defining equation $g^2=1-g$ to reduce all
powers of $g$ to first order. When we apply this reduction to $g^n$,
we obtain the remarkably simple formula $g^n=(-1)^n(F_{n-1}-gF_n)$, where
$F_n$ is the $n$th Fibonacci number (defined by $F_n=F_{n-1}+F_{n-2}$
for $n>2$, with $F_1=F_2=1$). For the golden walk, we now use this
construction to reduce the location of each endpoint, which is of the form
$\sum_n^N \epsilon_ng^n$, to the much simpler form $A+Bg$, where $A$ and $B$
are integers whose values depend on the walk. By this approach, each
endpoint location is obtained with perfect accuracy. The resulting
distribution, based on enumerating the exact probability distribution for
$N\leq 29$, is shown in Fig.~\ref{golden} at various spatial resolutions. At
$N=29$ this distribution is exact to a resolution of $10^{-7}$.

\subsection{Self-Similarity}

\begin{figure}[ht] 
 \vspace*{0.cm}
 \includegraphics*[width=0.35\textwidth]{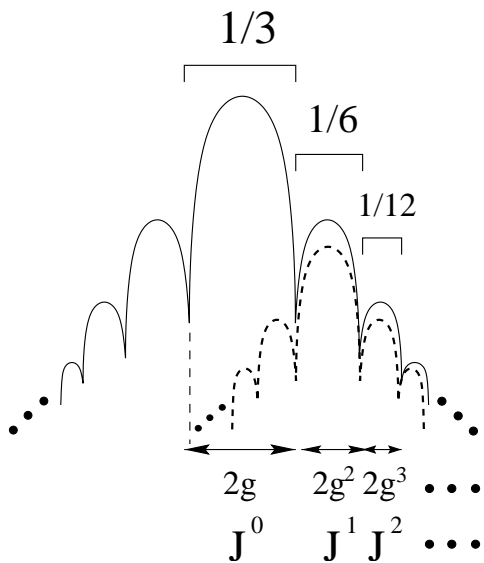}
\caption{Sketch to illustrate the symmetry and self-similarity of $P_g(x)$.
  The dashed curve is the probability distribution when the first step is to
  the right.  The full probability distribution is the sum of the dashed
  curve and an identical (but shifted) curve that stems from the distribution
  when the first step is to the left.  The measures associated with each lobe
  of $P_\lambda(x)$ (top) and the spatial extent of each lobe (bottom) are
  indicated.  Notice that the left extreme of the restricted distribution
  coincides with the first minimum of the full distribution.
\label{gold-scale}}
\end{figure}

Perhaps the most striking feature of the endpoint distribution is its
self-similarity, as sketched in Fig.~\ref{gold-scale}. Notice, for example,
that the portion of the distribution within the zeroth subinterval
$J^0=[-g,g]$ is a microcosm of the complete distribution in the entire
interval $J= [-g^{-2},g^{-2}]$. In fact, we shall see that the distribution
within $J^0$ reproduces the full distribution after rescaling the length by a
factor $g^{-3}$ and the probability by a factor of 3. Similarly, the
distribution in the first subinterval $J^1=[g,1+g^2]$ reproduces the full
distribution after translation to the left by 1, rescaling the length by
$g^{-4}$, and rescaling the probability by 6. A similar construction applies
for general subintervals.

To develop this self-similarity, it is instructive to construct the
symmetries of the probability distribution. Obviously, $P_g(x)$ is an even
function of $x$. That is,
\begin{equation}
\label{sym}
P_g(x)=P_g(-x).
\end{equation}
In fact, there is an infinite sequence of higher-order symmetries that arise
from the evenness of $P_g(x)$ about the endpoints after 1 step, 2 steps, 3
steps, etc.

For example, the first higher-order symmetry is
\begin{equation}
\label{sym1}
P_g(1+x)=P_g(1-x)
\end{equation}
for $|x|<g^2$. Equation~(\ref{sym1}) expresses the symmetry of $P_g(x)$ about
$x=1$ for the subset of walks whose first step is to the right. We can
ignore walks whose first step is to the left because the rightmost position
of such walks is
$-1+g+g^2+\ldots=g=1-g^2$. Thus within a distance of $g^2$ from $x=1$, only
walks with the first step to the right contribute to the distribution within
this restricted range. The probability distribution must therefore be
symmetric about $x=1$ within this same range.

\noindent
 \includegraphics*[width=0.42\textwidth]{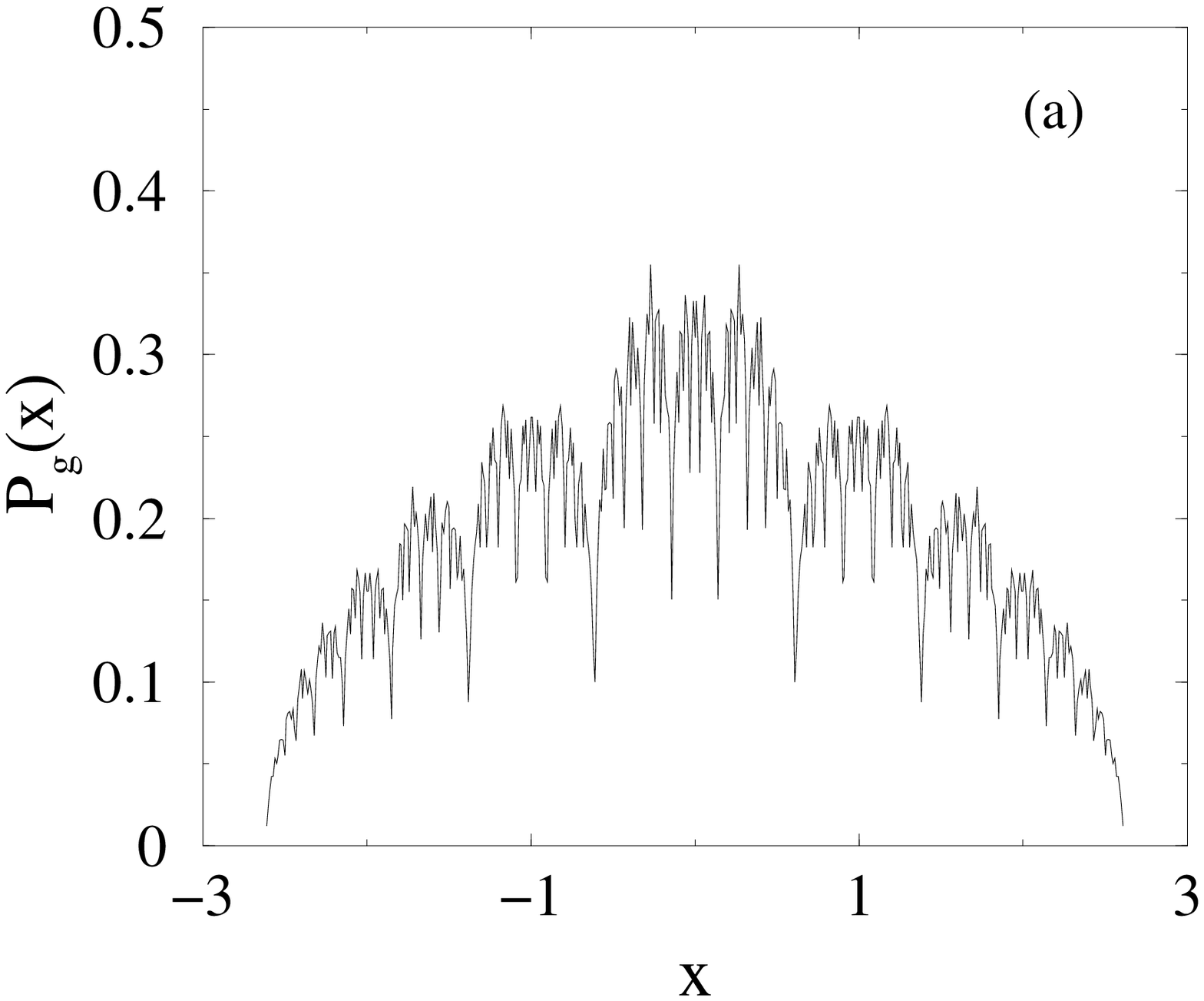}\\
 \includegraphics*[width=0.42\textwidth]{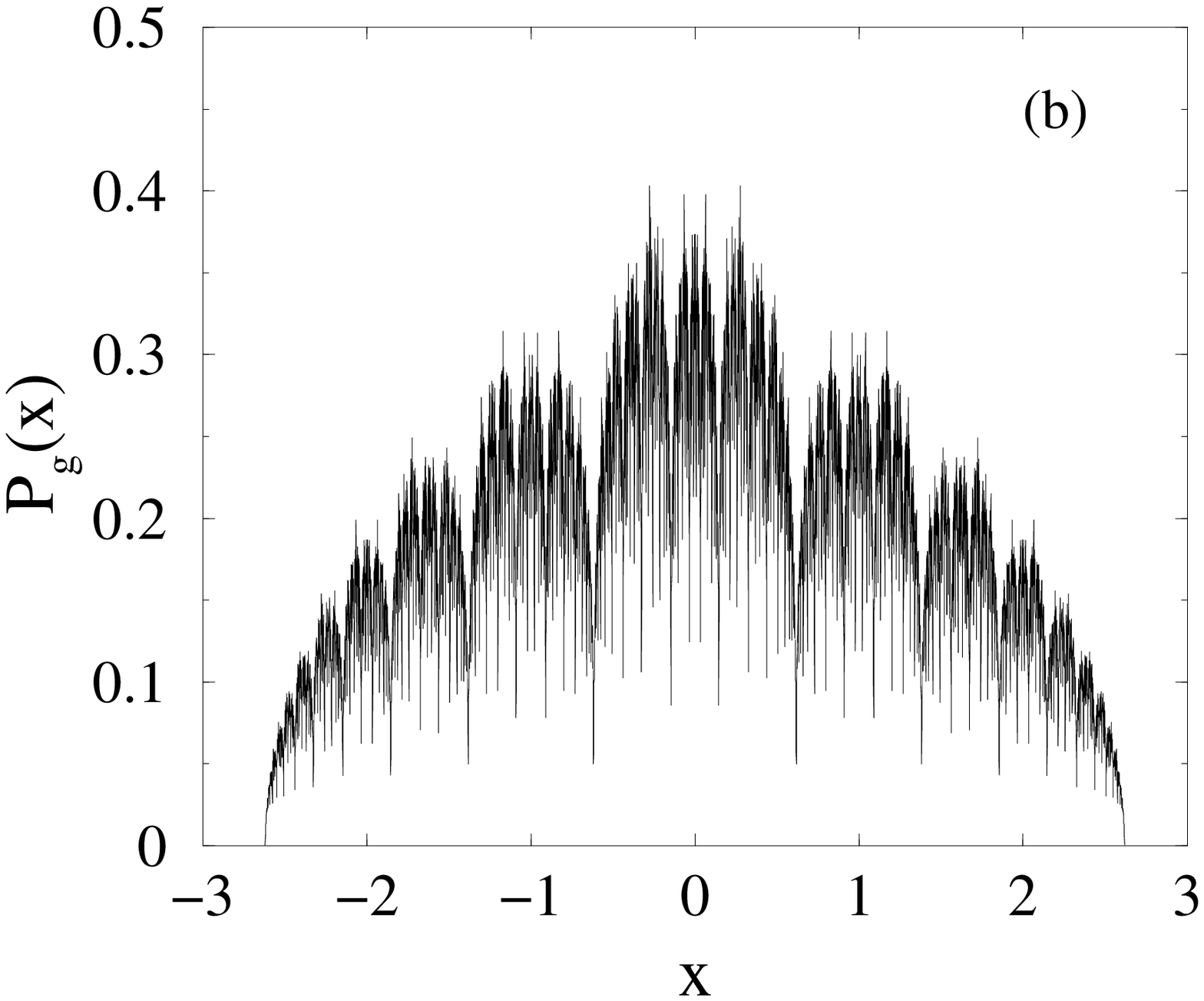}\\
 \includegraphics*[width=0.42\textwidth]{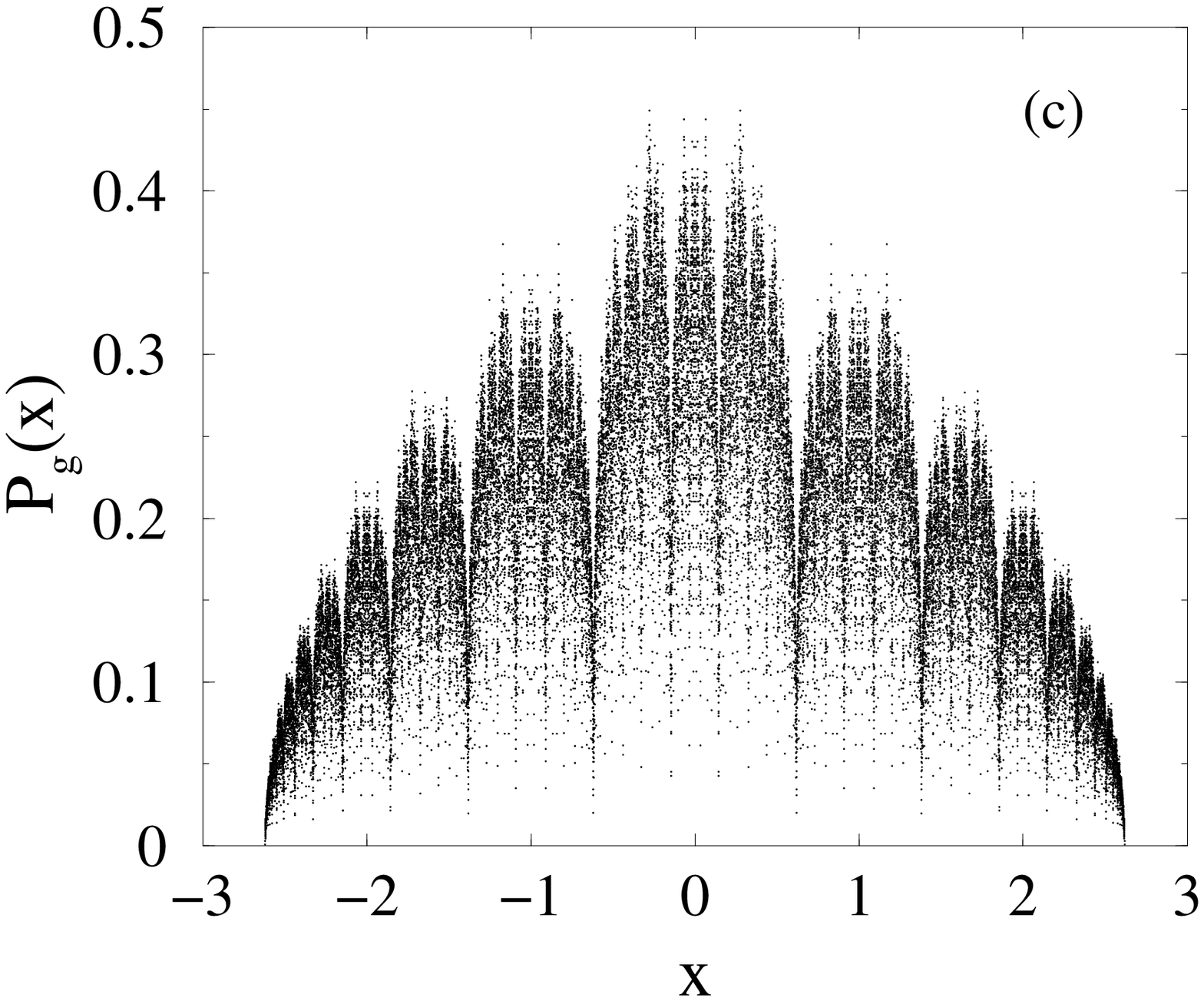}\\
\begin{figure}[H] 
 \vspace*{0.cm}
\caption{Probability distribution of the golden walk for a 29-step
  enumeration at spatial resolution $10^{-2}$, $10^{-3}$ and $10^{-4}$ (a) --
  (c), respectively.  In (c), the line joining successive points is not shown
  so that details of the distribution remain visible.
\label{golden}}
\end{figure}

Continuing this construction, there are infinitely many symmetries of the
form
\begin{equation}
\label{symk}
P_g\big(\sum_{n=0}^kg^n+x\big)=P_g\big(\sum_{n=0}^kg^n-x\big),
\end{equation}
with $k=1,2,\ldots$ that represent reflection symmetry about the point that
is reached when the first $k$ steps are all in one direction. The $k$th
symmetry applies within the range $|x|<g^{k+1}$. 

We now exploit these symmetries to obtain a simple picture for the measure of
the probability distribution, $M_g$. We start by decomposing the full
support $J$ into the contiguous subintervals that span the successive lobes
of the distribution, as shown in Fig.~\ref{gold-scale}. We label these
subintervals as $J^0=(-g,g)$, $J^1=(1-g^2,1+g^2)$, $J^2=(1+g-g^3,1+g+g^3)$,
etc.; there are also mirror image intervals to the left of the origin,
$J^{-k}=-J^k$.

We now use the invariance condition of Eq.~(\ref{M}) to determine the
measures of these fundamental intervals $J^k$. For $J^0=(-g,g)$, this
invariance condition yields
\begin{eqnarray}
M_g(-g,g)&=&\frac{1}{2}[M_g(-(2+g),-g)+ M_g(g, 2+g)]\nonumber \\
&=& \frac{1}{2}[1-M_g(-g,g)],
\end{eqnarray}
where the second line follows because of left-right symmetry and because the
intervals $(-(2+g),-g)$, $(g, 2+g)$, and $(-g,g)$ comprise the entire support
of a normalized distribution. We therefore obtain the remarkably simple
result that the measure of the central interval is $M_g(-g,g)=1/3$.
If we apply the same invariance condition to $J^1$, we find
$M_g (J^1)=\frac{1}{2} M_g(J^0)$. By continuing this same
reasoning to higher-order intervals $J^K$, we find, in general, that the
measure of the $k$th interval is one-half that of the previous
interval (see Fig.~\ref{gold-scale}). Thus we obtain the striking result
\begin{equation}
\label{mesk}
M_g (J^k)=\frac{1}{3\cdot 2^{|k|}}.
\end{equation}

\subsection{Singularities}

Another intriguing feature of $P_g(x)$ is the existence of a series of deep
minima in the distribution. Consider, for example, the most prominent minima
at $x=\pm g$ (see Fig.~\ref{golden}). The mechanism for these minima is
essentially the same reason that $g$ is sometimes termed the ``most
irrational'' real number -- that is, most difficult to approximate by a
rational number.\cite{numbers} In fact, there is only a single trajectory
of the random walk in which the endpoint of the walk reaches $g$, namely, the
trajectory that consists of alternating the steps, $1-g+g^2-g^3+g^4-\ldots$.
If there is any deviation from this alternating pattern, the endpoint of the
walk will necessarily be a finite distance away from $x=g$. This dearth of
trajectories with endpoints close to $x=g$ is responsible for the sharp
minimum in the probability distribution.

More generally, this same mechanism underlies each of the minima in the
distribution, including the singularity as $x\to x_{\max}$. For each
local minimum in the distribution, the first $n$ steps of the walk must be
prescribed for the endpoint to be within a distance of the order of $g^n$ to
the singularity. However, specifying the first $n$ steps means that the
probability for such walks can be no greater than $2^{-n}$. It is this
reduction factor that leads to all the minima in the distribution.

For simplicity, we focus on the extreme point in the following; the argument
for all the singularities is similar. If the first $n$ steps are to the
right, then the maximum distance $\Delta$ between the endpoint of the walk
and $x_{\max}$ arises if the remaining steps are all to the left. Therefore
\begin{eqnarray*}
\label{min}
\Delta=x_{\max}-(1+g+\ldots + g^n)+g^{n+1}+g^{n+2} + \ldots 
= 2g^{n-1}.
\end{eqnarray*}
Correspondingly, the total probability to have a random walk whose endpoint
is within this range is simply $2^{-n}$.

For $x$ near $x_{\max}$, we make the fundamental assumption that
$P_\lambda(x) \sim (x_{\max}-x)^\mu$. Although this hypothesis appears
difficult to justify rigorously for general values of $\lambda$, such a power
law behavior arises for $\lambda=2^{-1/m}$, as discussed in
Sec.~III. We assume that power-law behavior continues to hold for
general $\lambda$ values. With this assumption, the measure for a
random walk to be within the range $\Delta=x_{\max}-x$ of $x_{\max}$ is
$M(\Delta)\sim \Delta^{1+\mu}$. However, because such walks have the first
$n$ steps to the right, $M(\Delta)$ also equals $2^{-n}$. If we write
$\ln M = -n\ln 2$,
$\ln \Delta = +(n-1)\ln g + \ln 2$,
and eliminate $n$ from these relations, we obtain $M(\Delta)\sim \Delta^{\ln
2/\ln(1/g)}$ or, finally,
\begin{equation}
\label{tail}
P_g(\Delta)\sim \Delta^{-1+\ln 2/\ln(1/g)}.
\end{equation}
This power law also occurs at each of the singular points of the distribution
because the underlying mechanism is the same as that for the extreme points.

The same reasoning applies {\em mutatis mutandis} near the extreme points for
general $\lambda$, leading to the asymptotic behavior $P_\lambda(\Delta)\sim
\Delta^{-1+\ln 2/\ln(1/\lambda)}$. In particular, this reason gives, for the
tail of $\Pi_{m}$, the limiting behavior $\Delta^{m-1}$, in agreement with
the exact calculation in Sec.~III.

\section{Discussion}

We have outlined a number of appealing properties of random walks with
geometrically shrinking steps, in which the length of the $n$th step
equals $\lambda^n$ with $\lambda<1$. A very striking feature is that the
probability distribution of this walk depends sensitively on the shrinkage
factor $\lambda$ and much effort has been devoted to quantifying the
probability distribution. We worked out the probability distribution for the
special cases $\lambda=2^{-1/m}$ where a solution is possible by elementary
means. We also highlighted the beautiful self-similarity of the probability
distribution when $\lambda=(\sqrt{5}-1)/2$. Here, the unique features of
this number facilitate a numerically exact enumeration method and also lead
to very simple results for the probability measure.

We close with some suggestions for future work. What is the effect of a bias
on the limiting probability distribution of the walk? For example, suppose
that steps to the left and right occur independently and with probabilities
$p$ and $1-p$ respectively. It has been proven\cite{PS98} that the
probability distribution is singular for
$\lambda<p^p(1-p)^{1-p}$ and is continuous for almost all larger values of
$\lambda$. This is the analog of the transition at $\lambda=1/2$ for the
isotropic case. What other mysteries lurk within the anisotropic system?

Are there interesting first-passage characteristics? For example, what
is the probability that a walk, whose first step is to the right, never
enters the region $x<0$ by the $n$th step? Such questions are of
fundamental importance in the classical theory of random walks,\cite{weiss}
and it may prove fruitful to extend these considerations to geometric
walks. Clearly, for $\lambda <1$, this survival probability will approach
a non-zero value as the number of steps $N\to\infty$. How does the
survival probability converge to this limiting behavior as a function of
the number of steps? Are there qualitative changes in behavior as
$\lambda$ is varied?

What happens in higher spatial dimensions? This extension was suggested to us
by M. Bazant.\cite{bazant1} There are two natural alternatives that appear to
be unexplored. One natural way to construct the geometric random walk in
higher dimensions is to allow the direction of each step to be isotropically
distributed, but with the length of the $n$th step again equal to
$\lambda^n$. Clearly, if $\lambda \ll 1$, the probability distribution is
concentrated within a spherical shell of radius 1 and thickness of the order
of $\lambda/(1-\lambda)$. As $\lambda$ is increased, the probability
distribution eventually develops a peak near the origin.\cite{bazant} What is
the nature of this qualitative change in the probability distribution?
Another possibility\cite{brad} is to require that the steps are always
aligned along the coordinate axes. Then for sufficiently small $\lambda$ the
support of the walk would consist of a disconnected set, while as $\lambda$
increases beyond 1/2, a sequence of transitions similar to those found in one
dimension may arise.

\begin{acknowledgments}
We gratefully acknowledge NSF grants DMR9978902 and DMR0227670 for partial
support of this work. We thank Martin Bazant, Bill Bradley, and Jaehyuk Choi
for a stimulating discussion and advice on this problem. We also thank Boris
Solomyak for helpful comments on the manuscript.
\end{acknowledgments}

\appendix

\section{Fourier Inversion of the Probability Distribution for 
$\lambda=2^{-1/2}$}

For $\lambda=2^{-1/2}$, we write $P_\lambda(k)$ in the form
\begin{eqnarray}
\Pi_{2}(k) &=& \frac{\sin(2k)\sin(\sqrt{2}k)}{2^{3/2}k^2} \nonumber \\
&=& -\frac{(e^{2ik}-e^{-2ik})(e^{\sqrt{2}ik}-e^{-\sqrt{2}2ik})}{2^{7/2} k^2}
\nonumber \\
&\equiv& - \frac{[e^{ikx_2}+e^{-ikx_2}-e^{ikx_1}-e^{-ikx_1}]}{{2^{7/2}k^2}},
\end{eqnarray}
where $x_1=2-\sqrt{2}$ and $x_2=2+\sqrt{2}$. The inverse Fourier
transform is
\begin{equation}
\label{F}
\Pi_{2}(x)=\frac{1}{2\pi}\!\int_{-\infty}^\infty\! \Pi_{2}(k)\,
e^{-ikx}\, dk.
\end{equation}
To evaluate the integral, we extend it into the complex plane by including a
semi-circle at infinity. The outcome of this inverse transform depends on
the relation between $x$ and $x_1$, $x_2$.

For $x>x_2$, we must close the contour in the lower half-plane for each term,
so that the semi-circle contribution is zero. We must also indent the
contour around an infinitesimal semi-circle about the origin to avoid the
singularity at $k=0$. If $x>x_2$, the residues associated
with each term in the integrand cancel, and we obtain $\Pi_{2}(x)=0$.

For $0<x<x_1$, we must close the contours in the upper half-plane for the
first and third terms, and in the lower-half plane for the complementary
terms. The contribution of the first integral is proportional to
\begin{eqnarray*}
\oint \frac{e^{ik(x_2-x)}}{k^2}\,dk =i \pi \mbox{Res}
\Big[\frac{e^{ik(x_2-x)}}{k^2}\Big]\Bigg|_{k=0} 
= -\pi(x_2-x).
\end{eqnarray*}
Similarly, the contributions of the remaining three integrals are
$-\pi(x_2+x)$, $-\pi(x_1-x)$,and $-\pi(x_1+x)$, respectively. As a result,
we find, for $0<x<x_1$, 
\begin{eqnarray*}
\Pi_{2}(x)=\frac{(x_2 - x) + (x_2 + x)
- (x_1 - x) - (x_1 + x)}{2^{9/2}}
= \frac{1}{4}. 
\end{eqnarray*}
Finally, for $x_1<x<x_2$, we must close the contour in the upper half-plane
for the first two terms in Eq.~(\ref{F}) and in the lower-half plane for the
latter two terms. Evaluating each of the residues, we now obtain, for $x_1<x<x_2$
\begin{eqnarray*}
\Pi_{2}(x)=\frac{1}{4}\big(1-\frac{x}{x_2}\big). 
\end{eqnarray*}

\newpage

\end{document}